\documentclass{osa-article}

\journal{oe}

\articletype{Research Article}
\usepackage{braket}
\usepackage{upgreek}
\usepackage{todonotes}

\begin{document}

\title{Curved GaAs cantilever waveguides for the vertical coupling to photonic integrated circuits}
\author{Celeste Qvotrup,\authormark{1,*} Zhe Liu,\authormark{1} Camille Papon,\authormark{1}, Andreas D. Wieck,\authormark{2}  Arne Ludwig,\authormark{2} and Leonardo Midolo\authormark{1}}

\address{\authormark{1}Center for Hybrid Quantum Networks (Hy-Q), Niels Bohr Institute, University of Copenhagen, Blegdamsvej 17, DK-2100 Copenhagen, Denmark. \\
\authormark{2}Lehrstuhl f\"ur Angewandte Festk\"orperphysik, Ruhr-Universit\"at Bochum,
Universit\"atsstrasse 150, D-44780 Bochum, Germany.}

\email{\authormark{*}celeste.qvotrup@nbi.ku.dk} 

\begin{abstract}
We report the nanofabrication and characterization of optical spot-size converters couplers based on curved GaAs cantilever waveguides.  Using the stress mismatch between the GaAs substrate and deposited Cr-Ni-Au strips, single-mode waveguides can be bent out-of-plane in a controllable manner. A stable and vertical orientation of the out-coupler is achieved by locking the spot-size converter at a fixed 90$^\circ$ angle via short-range forces. The optical transmission is characterized as a function of temperature and polarization, resulting in a broad-band chip-to-fiber coupling extending over a 200 nm wavelength bandwidth. The methods reported here are fully compatible with quantum photonic integrated circuit technology with quantum dot emitters, and open opportunities to design novel photonic devices with enhanced functionality.
\end{abstract}

\section{Introduction}
Epitaxially-grown gallium arsenide (GaAs) membranes are a promising technology for the development of quantum photonic integrated circuits (QPICs) featuring waveguide-integrated single-photon emitters, routers, and detectors \cite{dietrich2016gaas,hepp2019semiconductor,uppu2021quantum}. 
The use of suspended membranes enables two-dimensional light confinement in the chip plane owing to the large refractive index contrast between semiconductor and vacuum, and enhances the coupling efficiency between quantum emitters, such as self-assembled InAs Stranski-Krastanov \cite{marzin1994photoluminescence} or GaAs droplet \cite{huber2017highly} quantum dots (QDs), and waveguide modes to near-unity values \cite{arcari2014near}. 

To reliably characterize and operate QPICs, a method for the broadband and efficient coupling of light in waveguides is essential. 
Additionally, it is often necessary to optically access the chip from the surface, rather than from the edge, for example when testing several devices across the chip, or when exciting quantum emitters. A wide variety of strategies for coupling light into PICs have been developed, primarily targeting the silicon photonics platforms \cite{Marchetti:19}.
Grating couplers \cite{taillaert2004compact,zhou2018high} are commonly employed for direct fiber-coupling or to focus light beams with microscope objectives, but typically offer a limited bandwidth (<30 nm) and are highly polarization-sensitive. 
On the contrary, edge coupling methods, such as inverted tapers with spot-size converters based on evanescent coupling \cite{roelkens2005efficient}, offer high efficiency, large spectral bandwidth, and polarization independence at the cost of requiring optical access from the side, and thus restricting the positioning of the couplers to the edge of the chip. 
An ideal broadband coupler could be created by implementing spot-size converters tilted out of the chip plane. 

\begin{figure}[htbp!]
\centering\includegraphics[width=13cm]{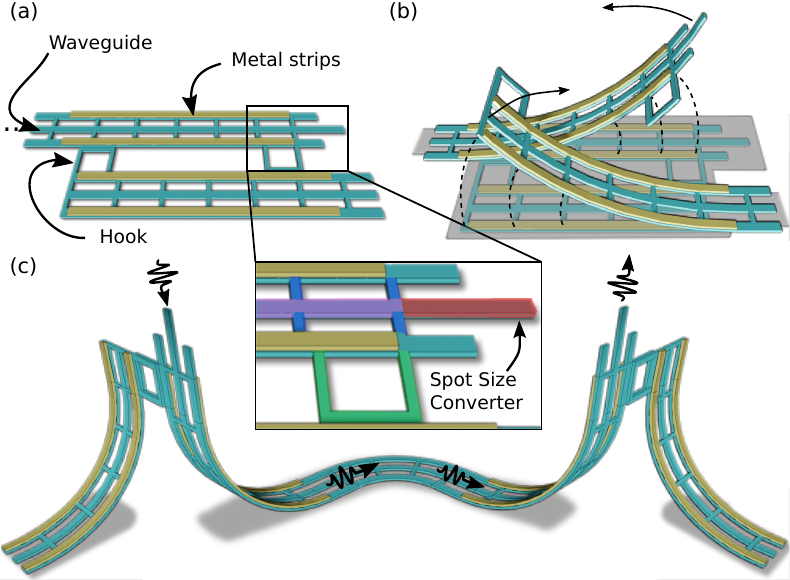}
\caption{\label{fig:fig1}  Curved cantilever waveguide couplers and locking mechanism. (a)  Illustration of the waveguide before being released from the substrate. The upper waveguide connects to the planar circuit and is terminated with a spot-size converter at the tip. The lower structure provides a mechanical counter-weight to lock the waveguides vertically. (b) Once the structures are released from the substrate, the cantilevers curl up to minimize intrinsic stress mismatch. Two protruding ``hooks'' stabilize the coupler at a vertical position. (c) Sketch of a transmission device consisting of two connected bent cantilever couplers. Light is coupled in through one tip, travels through the planar waveguide circuit, and couples out through the other tip. Inset: detail of the spot-size converter (red), waveguide (magenta), suspension tethers (blue), and hook (green).}
\end{figure}

In this work we demonstrate a coupler created by bending a free-standing GaAs cantilever waveguide out of the chip plane with a polymer spot-size converter \cite{uugurlu2020suspended} fabricated on its tip. The bending is achieved by depositing thin strips of nickel, which provide a large tensile strain that seeks to relax when cantilevers are undercut, causing nanostructures to ``roll-up'' when released. Unlike previous demonstrations of curved cantilevers \cite{Yoshida:15,Sun:09}, where no control was demonstrated on the output angle, we show how a nearly-perfect 90 degree orientation can be obtained by locking the waveguides into a fixed position (Fig.~\ref{fig:fig1}). The lock mechanism, based on short-range inter-solid adhesion forces, maintains the vertical orientation after cooling down the sample to cryogenic temperatures, as required for the operation of quantum emitters and single-photon detectors. The couplers offer a broad band coupling extending from 870 nm up to 1050 nm, close to the 1070 nm upper limit of the utilized spectrometer,  at a temperature of 10 K. The fabrication method presented here opens up avenues for realizing three-dimensional folding of nanophotonic structures comprising waveguides and quantum emitters. It further offers a versatile approach to overcome stiction in nano-opto-electro-mechanical systems \cite{midolo2018nano}, by engineering the stress locally on fragile nanostructures, preventing their collapse after membrane undercut.

\section{Device concept and working principle}
Figure \ref{fig:fig1} shows an artistic impression of the curved cantilever waveguide couplers developed in this work and the mechanism enabling a fixed 90-degree orientation. The waveguides are patterned on GaAs with a periodic set of thin tethers on each side, connecting them rigidly to two bilayer GaAs/metal cantilevers, as shown in Fig.~\ref{fig:fig1}(a). The tethers, which are designed to reduce optical scattering, allow distancing the metal from the waveguides to avoid optical loss. When released from the substrate, the cantilever waveguides bend upwards as shown in  Fig.~\ref{fig:fig1}(b).

Previous works have demonstrated curved waveguides in the silicon-on-insulator platform by using either ion implantation \cite{Yoshida:15}, or evaporated Ti/Ni layers followed by high-temperature rapid thermal annealing to control the radius of curvature \cite{Sun:09, Sun:11}. In GaAs, a large stress is conventionally achieved by embedding lattice-mismatched layers, e.g. quantum wells, in the epitaxial growth of heterostructures that cause roll-up after undercut \cite{Grimm:13}. However, this method does not allow to apply the stress locally, i.e. to a specific part of the chip.
Here, we use unstrained lattice-matched GaAs membranes epitaxially grown on a Al$_{0.75}$Ga$_{0.25}$As sacrificial layer, and provide the required stress by depositing strips of nickel (Ni) by electron-beam evaporation, which is known to provide a large intrinsic stress during deposition, which depends on process chamber temperature and evaporation rate \cite{Moiseeva2007SinglemaskMO}.

To achieve a fixed 90-degree orientation, orthogonal to the chip plane, two nearly-identical cantilevers are patterned next to each other, so that they roll up in opposite directions upon release. Two protruding ``hook'' structures are patterned close to the tips of both cantilevers (indicated by arrows in Fig.~\ref{fig:fig1}(a)) so that they touch when the bending reaches a 90-degree angle. Owing to the symmetry of the two cantilevers, the bending forces counteract each other preventing them from bending further, as shown in Fig.~\ref{fig:fig1}(c), while short-range electrostatic forces keep the two hooks connected and prevent them from moving apart upon changes in temperature.  
Using this method, it is possible to lift and lock at 90-degree orientation virtually any nanostructure fabricated on the tip of the cantilever. To demonstrate this concept on a practical device, we have fabricated spot-size converters using the procedure outlined in Ref. \cite{uugurlu2020suspended} to realize a vertical optical coupler. Figure \ref{fig:fig1}(c) illustrates two of such couplers connected via a planar waveguide, enabling the adiabatic coupling of light in and out of the chip plane. 

\section{Device fabrication}
\begin{figure}[htbp]
\centering\includegraphics[width=13cm]{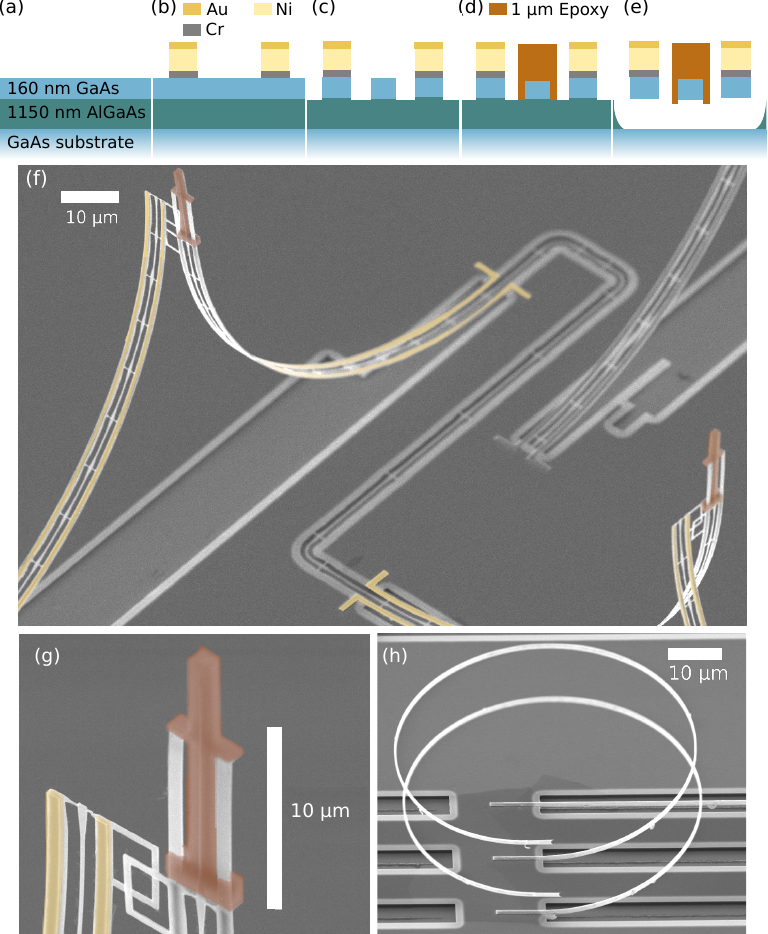}
\caption{\label{fig:fig2} Fabrication procedure and results of the bent cantilever waveguide couplers. (a) Wafer layout. (b) Cr/Ni/Au metal strip evaporation. (c) Reactive ion etching (RIE) of GaAs waveguides and cantilevers. (d) Definition of overlay polymer via negative epoxy-based resist.  (e) Undercut in HF solution. (f) False color SEM image of a bent cantilever waveguide device. The epoxy overlay polymer is highlighted in brown, nickel strips in yellow. (g) Detail of the spot-size converter.  (h) Proof-of-concept demonstration of 360 degree bending on a bi-layer cantilever. }
\end{figure}

The bent cantilever waveguide devices are fabricated on an undoped GaAs membrane grown on Al$_{0.75}$Ga$_{0.25}$As, as shown in figure \ref{fig:fig2}(a). Initially, metal strips for the bilayer cantilevers as well as alignment markers are patterned using electron beam (e-beam) lithography on a 550 nm thick positive tone resist (CSAR 13) followed by 10 nm Cr, 160 nm Ni and 10 nm Au being deposited by Electron Beam Physical Vapour Deposition, as can be seen in figure \ref{fig:fig2}(b). Here, the thin Cr layer is deposited to improve the metal adhesion, at the expense of a slightly reduced stress, while the Au layer prevents surface oxidation.
Cantilevers and suspended waveguides are defined using e-beam exposure after spin coating with a 200 nm thick positive tone resist (CSAR9) on top of an adhesion promoter (OmniCoat) followed by reactive ion etching into the GaAs membrane in a BCl$_3$/Cl$_2$/Ar plasma (illustrated in figure \ref{fig:fig2}(c)). The etching process is stopped a few tens of nm inside the AlGaAs sacrificial layer. 

The waveguides are terminated with a 15-$\upmu$m-long linear inverse taper to a 60-nm-wide tip. To fabricate the spot-size converters, a negative epoxy resist (EpoCore 5, Microresist GmbH) is spin-coated to a thickness of 1150 nm and a 20-$\upmu$m-long overlay waveguide is exposed via electron-beam lithography, covering the GaAs taper \cite{uugurlu2020suspended}, as shown in figure \ref{fig:fig2}(d). Finally, the sample is undercut in a 5\% hydrofluoric acid solution, rinsed, and dried via supercritical CO$_2$ drying as in figure \ref{fig:fig2}(e).
During the undercut process, the cantilevers are released and the intrinsic strain in the bilayer cantilever relaxes by curling the waveguide out of plane.

\subsection{Characterization of the bending radius}
\begin{figure}[htbp]
\centering\includegraphics[width=13cm]{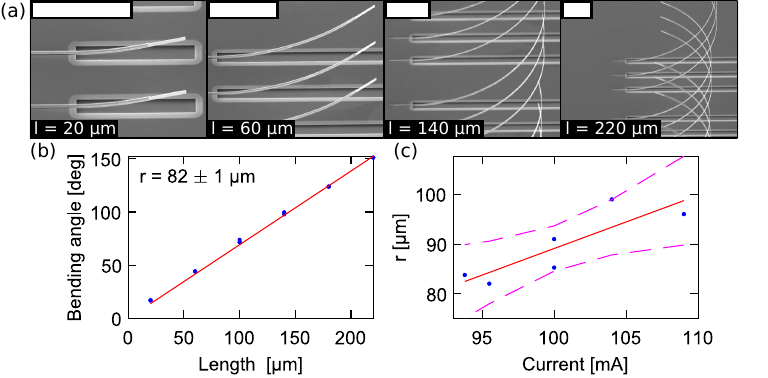}
\caption{\label{fig:fig3} Bending radius measurements. (a) SEM images of several cantilevers with varying metal strip lengths. For all the images, the white scalebar corresponds to 25 $\upmu$m. (b) Output angle with respect to the wafer plane as a function of the bi-layer strip length for a Ni deposition rate of 2 \AA/s. The slope of the linear fit is the inverse of the curvature $r$. (c) Bending radii as a function of the e-beam evaporation current with constant Ni deposition rate at 2 \AA/s. The purple lines show the fit confidence bounds. }
\end{figure}

For a bi-layer cantilever made of two layers of identical thickness $t=160$ nm, containing intrinsic strain mismatch, the radius of curvature $r$ after release is given by \cite{Moiseeva2007SinglemaskMO} 
\begin{equation}
r=\frac{t}{\Delta\epsilon}\left(\frac{(\eta +1)^2}{12\eta}+1\right),
\label{eq:brad}
\end{equation} 
where $\eta=E_{\textrm{Ni}}/E_{\textrm{GaAs}}$ is the ratio between the Young's modulus of the two layers, and $\Delta\epsilon$ is the misfit strain defined as the strain difference between the Ni and GaAs layer after deposition. Here, we assume that the GaAs is un-strained, owing to lattice-matched epitaxial growth, while the evaporated Ni layer possesses a compressive bi-axial stress $\sigma_\textrm{Ni}$ leading to a strain $\Delta\epsilon = (1-\nu)\sigma_\textrm{Ni}/E_{\textrm{Ni}}$, where $\nu=0.31$ is the Poisson ratio of nickel. 
To estimate the stress in our deposited films, we fabricated several cantilevers with lengths varying from 20 to 220 $\upmu$m, whose scanning electron microscope (SEM) images are shown in Fig.~\ref{fig:fig3}(a). From the analysis of SEM images, we determine the angle $\theta = L/r$ formed by the released cantilevers with respect to the substrate plane as a function of the cantilever length L. The result is shown in Fig.~\ref{fig:fig3}(b), from which a bending radius $r=(82 \pm 1) \:\upmu$m is derived. The linear dependence indicates that the bending radius is highly reproducible across many different devices, enabling consistent results within the same processing run.

From Eq. \ref{eq:brad}, we derive a total strain of $\Delta\epsilon \simeq 2.7\times10^{-3}$, corresponding to a bi-axial stress of roughly 860 MPa, compatible with the values reported in the literature, given at 800--900 MPa for similar evaporation rates of 2 \AA/s \cite{strifler1991stress,mihailovich1996measurement}. The total stress is the result of two contributions, i.e. the internal stress of the film and the thermal stress caused by heating of the sample during the depositon, the latter of which is partially caused by the difference of thermal expansion coefficients between the two materials, and partially by the difference in temperature between the film and the substrate during Ni deposition. The latter can vary substantially depending on the deposition rate and the overall status, geometry and configuration of the e-beam evaporator chamber, resulting in different bending radii at each evaporation run. Carrying out several evaporation tests, a weak correlation was found between variation of the bending radius and the fluctuations of the e-beam current that controls the evaporation rate of nickel in the deposition tool, up to approximately $\pm 10\: \upmu$m for a 5\% fluctuation in the current.
The bounds shown in Fig. \ref{fig:fig3}(c) by dashed purple lines indicate the interval of confidence of the linear fit to the data, showing that a precise control of the final bending radius below $10 \upmu$m cannot be achieved with this method. 
Moreover, operating the devices at different temperatures, for example under cryogenic conditions (4 K) as often required in quantum applications, will cause further bending due to thermal stresses. It is therefore desirable to ``lock'' the waveguides in a fixed vertical position after they have been released from the substrate.

\subsection{Design of the lock mechanism}
\begin{figure}[htbp]
\centering\includegraphics[width=13cm]{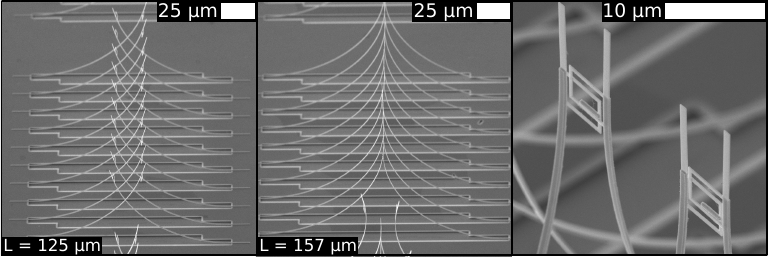}
\caption{\label{fig:fig4}Technique to achieve a stable vertical orientation of the cantilevers. (a) SEM image of ten cantilevers with length $L= 125$ $\upmu$m right after fabrication in an ``unlocked'' state. (b) Same as (a) but with longer cantilevers ($L=157$ $\upmu$m). (c) Detail of the hook microstructure used to achieve permanent locking. All images are taken at room temperature.}
\end{figure}

To achieve the locking, we prepared a set of cantilever pairs that roll-up in opposite directions. Both cantilevers feature a rectangular frame patterned on GaAs, which protrudes on one side. Such rectangular ``hooks'' are designed to collide with each other upon release. Taking advantage of short-range inter-solid adhesion forces (electrostatic, Van der Waals), these frames provide a fixed anchor to both cantilevers, ensuring a nearly-perfect vertical orientation.
We demonstrate this effect in Fig.~\ref{fig:fig4}(a) (\ref{fig:fig4}(b)), where ten cantilever pairs of length $L=125\: \upmu$m ($L = 157\: \upmu$m), are shown right after fabrication in an unlocked (locked) state.

In a bimorph cantilever, the thermal stress is given by \cite{strifler1991stress}
\begin{equation}
\sigma_\textrm{th} = \frac{E_\textrm{Ni}}{1-\nu}(\alpha_\textrm{GaAs}-\alpha_\textrm{Ni}) \Delta T,
\label{eq:therm_stress}
\end{equation} 
where $\alpha_\textrm{Ni} = 13.4 \times 10^{-6}$ 1/K and $\alpha_\textrm{GaAs} = 5.6 \times 10^{-6}$ 1/K are the thermal expansion coefficients of the two materials at room temperature and $\Delta T$ is the temperature difference. Both materials thus compress when cooling down, although Nickel does so at a double rate, resulting in an effective additional tensile strain which causes the bending radius to reduce further. Since a precise measurement of the bending radius as a function of temperature would require a cryogenic imaging system, we only provide an estimate of the added strain from equation \ref{eq:therm_stress}, to be in the order of $\sim 10^{-3}$. Using this final strain and equation \ref{eq:brad}, a final cryogenic bending radius of $\sim 70 \:\upmu$m can be derived. Thus, at low temperatures, unconnected cantilevers curl up further, causing the hook structures to collapse onto each other, while connected cantilevers will stay connected due to the forces exerted by the hook.
This method allows to either fabricate the structures with fixed orientation directly upon release, or in operando via cryogenic cooling. 

\section{Numerical simulation of bent cantilever waveguides}
To verify that the model of Eq. \ref{eq:brad} is consistent with more complex geometries, such as the one proposed in Fig. \ref{fig:fig1}, and with anisotropic materials such as GaAs, we perform a numerical simulation of the bent cantilever waveguides using finite-element analysis. The model consists of a flat cantilever waveguide, with a design identical to the one fabricated, oriented along the [110] direction of GaAs. We use the GaAs elastic compliance matrix \cite{hjort1994gallium} in a frame rotated by 45 degrees around the [001] growth axis, to account for the orientation of the waveguides in fabricated samples.

The model includes geometric non-linearities that take into account the out-of-plane rotation of the structure, enabling the correct estimation of the bending radius. A bi-axial stress is applied to the Nickel layer and it is oriented along the [100] and [010] directions, then gradually swept up to the maximum value of 925 MPa for a 122 $\upmu$m-long cantilever, until a vertical orientation of the tip is achieved. 

\begin{figure}[htbp]
\centering\includegraphics[width=13cm]{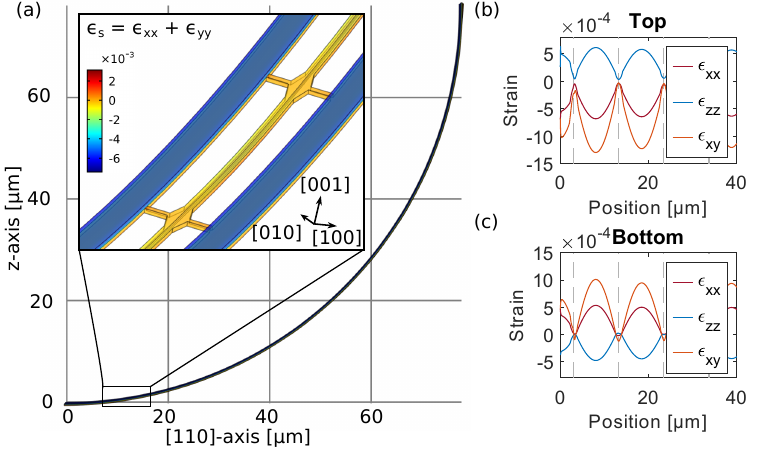}
\caption{\label{fig:fig5} (a): Numerical simulation of the bent cantilever waveguide with an intrinsic stress $\sigma_{xx}=\sigma_{yy}=925\:\text{MPa}$ applied to the nickel film. Inset: surface strain $\epsilon_{s} = \epsilon_{xx}+\epsilon_{yy}$ on a section of the waveguide between two consecutive tethers. (b) Non-zero strain components $\epsilon_{xx}=\epsilon_{yy}$, $\epsilon_{zz}$ and $\epsilon_{xy}$ at the top of the waveguide oriented along the $[110]$ crystal axis. (c) Same as (b), but for the bottom surface of the waveguide. The grey dashed lines show the position of tethers. }
\end{figure}

Figure \ref{fig:fig5}(a) shows the simulated deformation of the cantilever and the total strain induced on the waveguide (inset) between two tethers. The strain is defined as the uni-axial strain $\epsilon_s$, given by the sum of the two planar strain component $\epsilon_{xx}+\epsilon_{yy}$, where $x$ and $y$ refer to the GaAs crystallographic axes ([100] and [010]) in the material frame. On the central part of the waveguide a maximum strain $\epsilon_s \simeq 1\times10^{-3}$ is expected, with positive sign (tensile) at the bottom of the waveguide and negative sign (compressive) at the top surface.  
The spatial dependence of the non-zero strain components ($\epsilon_{xx} = \epsilon_{yy}$, $\epsilon_{zz}$, and the in-plane shear strain $\epsilon_{xy}$) along the waveguide length are shown in Fig.~\ref{fig:fig5}(b) and (c) for the top and the bottom of the waveguide, respectively. The strains are plotted over a 40 $\upmu$m-long section, showing a periodic pattern across the entire length of the cantilever, due to the presence of the support tethers every 10 $\upmu$m. 
Given the symmetry of the structure, the top and bottom strains are almost exactly flipped in sign, and cancel in the middle of the waveguide (neutral axis), suggesting that quantum emitters placed in the center of the waveguide will be largely unaffected by the bending.

\section{Characterization of transmission}
\begin{figure}[htbp]
\centering\includegraphics[width=13cm]{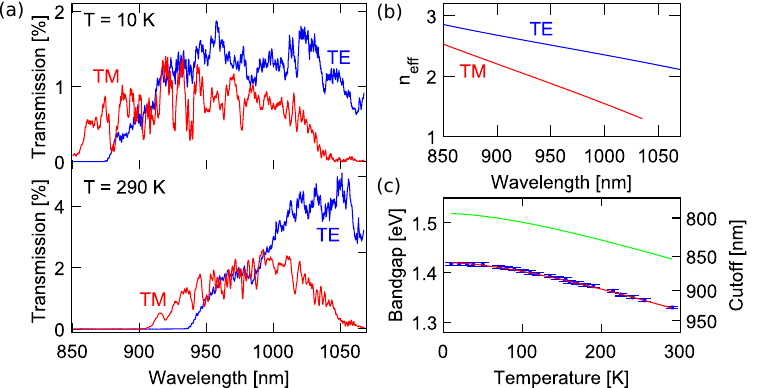}
\caption{\label{fig:fig6} (a) Transmission of TE (blue) and TM (red) polarized light across the bent waveguide carrying out the measurement shown in  Fig.~\ref{fig:fig1}(c) at 10 K and 290 K. (b) Simulation of the effective mode index for TE (blue) and TM (red) light across the measured wavelength range. (b) Measured cutoff transmission energy of the TE transmission as a function of temperatures (blue dots) and fit (red solid line). The temperature dependency of the GaAs bandgap is also shown (green solid line) for comparison.}
\end{figure}
The sample is tested in a He-flow cryostat with optical access through the top window, where transmission characterization measurements were carried out at room temperature, 10 K, and intermediate temperatures. 
The polarization of the light entering the circuit was controlled by fine-tuning the orientation of a set of half and quarter wave plates placed in the excitation path of the experiment. By coupling a continously tuneable laser into a set of focusing shallow etched gratings (SEGs)\cite{zhou2018high}, which facilitate transmission of the Transverse Electric (TE) but not the Transverse Magnetic (TM) mode, fabricated on another chip mounted on the PCB, the waveplate configuration resulting in the polarizations coupling to these modes were identified.
The efficiency, bandwidth and polarization dependence of the bent waveguide cantilever devices were then characterized using a supercontinuum laser (NKT SuperK Extreme).

One coupler of the device is excited via an objective focusing light on the plane of the cantilever tip (i.e. around 100 $\upmu$m above the chip surface). The light emerging from the output coupler is collected via the same objective and analyzed with a spectrometer covering a wavelength range of 850-1070 nm. 
The transmitted signal is shown in Fig. \ref{fig:fig6}(a) at respectively cryogenic ($T=10$ K) and room temperatures and for both TE and TM modes.

A large transmission bandwidth is observed for TE-polarized light, extending over 200 nm at 10 K and significantly wider than typical reported values for grating couplers, which are usually limited by the Bragg condition to $\ll 50$ nm full-width half-maximum. TM polarization has a slightly lower bandwidth due to the loss of mode confinement at longer wavelengths. To confirm this, Fig. \ref{fig:fig6}(b) shows the simulated effective mode index for a standard single-mode waveguide (width of 300 nm and thickness of 160 nm) as a function of wavelength for both TE and TM polarization at 10 K. While TE light maintains an effective refractive index higher than 2 for the entire experimental range, the TM mode is very weakly confined at $\lambda > 1000$ nm, resulting in a reduced transmission at longer wavelengths.

To obtain a calibration reference for the transmission and estimate the coupler efficiency, we first perform a confocal reflectivity measurement by focusing the laser to a homogeneous section of the GaAs surfacem \cite{zhou2018high}. The acquired spectrum provides a reference intensity $I_{\text{ref}}$, that includes the losses through the optical excitation paths. The recorded spectrum of the light transmitted across the whole device, $I_{\text{tr}}$, is then normalized to the reference signal reflected by the sample, i.e. $T_{\text{dev}} = (I_{\text{tr}}/I_{\text{ref}}R_{\text{GaAs}})$ where $R_\text{GaAs}\approx 0.31$ is the reflection coefficient for bulk GaAs at normal incidence (assuming a constant refractive index of 3.5 across the entire measurement spectrum). 

Using this method, we estimate the maximum total  transmission across the circuit for the TE mode to be $1.9\%$ at 958 nm and $5.1\%$ at 1051 nm for cryogenic and room temperature, respectively. One possible explanation for the lower transmission at cryogenic temperatures is the temperature drop causing additional strain mismatch between the GaAs and Ni layers, causing additional buckling of the device.
These efficiencies presented here are the result of the combined loss of the two input-output couplers and the propagation losses in the waveguide. In a previous work, we have measured accurately the loss per unit length for TE light at 10 K in undoped suspended waveguides, fabricated with the same procedure of this work, and obtained  $\alpha_\textrm{TE} = -7.5 \pm 1.0$ dB/mm at 930 nm. Given the total device length of $0.435$ mm, an attenuation of (-3.3 $\pm$ 0.4) dB due to propagation losses is expected. 
The maximum efficiency of a single coupler, assuming both couplers are identical, are therefore estimated is therefore estimated by $T_{\text{coupler}} = \sqrt{\frac{T_{\text{dev}}}{T_{\text{wg}}}}$ to be  $(20 \pm 1 )\% $ at 958 nm and $(32 \pm 2 )\% $ at 1051 nm for respectively cryogenic and room temperatures. At 930 nm, a common wavelength for InAs quantum emitters, the efficiency of a single coupler can be estimated to be $(15 \pm 2)$\%. This estimation is a lower bound, as it is likely that one of the couplers is not optimized for excitation or collection with the free-space optics setup utilized for the experiment. While the efficiency is lower than other methods reported in the literature, we believe it can be further improved by direct fiber coupling or by a more controlled focusing system, capable of matching the mode size in the spot-size converter tip.

\section{Discussion}
A remarkable feature of the transmission spectra across bent cantilever waveguides is the clear temperature- and polarization-dependent cut-off at shorter wavelengths. Such behavior is not observed in planar waveguides, suggesting that it originates as a consequence of the waveguide bending. Additionally, the cut-off does not vary with optical power, ruling out any potential thermal effect. In Fig. \ref{fig:fig6}(c), we plot the measured cut-off wavelength for TE light as a function of the sample temperature and compared it to the theoretical bandgap energy\cite{blakemore1982bandgap}. By fitting the data with the same model, we conclude that the cut-off is related to a shift of the GaAs bandgap caused by the curved cantilevers. 
It has been previously reported in several works that a moderate uni-axial strain can produce a large bandgap shift \cite{signorello2013tuning} and even flip the quantization axis of GaAs quantum dots \cite{yuan2018uniaxial}, which control the optical selection rules. In bent cantilevers, the top and bottom surfaces experience a large compressive and tensile uni-axial strain, respectively, with magnitude $|\epsilon_s| \simeq 0.1$\%, oriented along the direction of the waveguide (cf. Fig \ref{fig:fig5}(b) and (c)). 

As the waveguides are fabricated along the cleaving axes of GaAs, the in-plane strain in the material coordinate system is distributed bi-axially with an added shear component, i.e.  $\epsilon_{xx}=\epsilon_{yy}=\frac{\epsilon_s}{2}\simeq\pm\epsilon_{xy}$, where the plus sign is for waveguides along $[110]$ and the minus sign holds for waveguides along $[1\bar{1}0]$. 
According to the strain theory of semiconductor in the  $\mathbf{k}\cdot\mathbf{p}$ approximation \cite{sun2009strain}, bi-axial strain lifts the degeneracy between light and heavy holes in the valence band, while shear strain is responsible for mixing their orbitals. As a consequence, the dipolar matrix elements $\mathbf{d}_{vc}$ for transitions between conduction and valence bands are oriented differently on the surface of the cantilever than in flat, un-strained waveguides.
The exact determination of the bandgap shift from cut-off measurements and the consequent absorption strength is not trivial, as losses occur predominantly on the surface of the waveguide. Here, the combination of lattice disorder, mid-gap states, surface oxides, and stress produce Fermi-level pinning and piezoelectric-induced electric fields, which are likely to enhance the absorption below the gap over several tens of meV, e.g. due to electroabsorption \cite{wang2021electroabsorption}. Moreover, electroabsorption is highly polarization-dependent, which could also explain the difference in cut-off for the TE and TM modes, that couple differently to light- and heavy-hole valence bands. Discerning the individual contributions from all these effects is however beyond the scope of this work and requires further investigation. For the present demonstration, the cut-off  has little impact on the device performance at cryogenic temperatures and at 930 nm, where quantum dots are usually operated. Bandgap effects in strained GaAs remain, however, an interesting aspect, that future experiments will need to take into account. 

\section{Conclusions}
We have shown the design, fabrication and characterization of an out-of-plane coupler for integrated photonics utilizing a cantilever waveguide bent in a 90 degree angle due to strain mismatch between the GaAs membrane and an evaporated Ni film. This device exhibits broadband coupling for both TE and TM modes, with the TE mode of a single coupler having a 3-dB transmission bandwidth of >150 nm at cryogenic temperatures. Towards the development of robust  and more efficient chip-to-fiber couplers, future work will involve encapsulating the structures to gain mechanical stability and use direct fiber coupling instead of confocal setups. 

The fabrication method outlined in this work in combination with the accurate modeling provided by finite element simulations, provide a versatile tool to engineer strains in nanophotonic circuits and nanostructures, which is not only limited to GaAs material. Applications of such technique include the precise control of mechanical frequencies in resonators, the development of folding nano-structures, and preventing stiction upon wet etching due to capillary forces. The combination of strain engineering with quantum emitters is also a promising area to achieve control over the emission wavelength of the quantum dots, by placing them away from the neutral axis. Active control on the strain can also be achieved, for example using the Ni films as electrodes and exploiting piezoelectricity in GaAs. Finally, the large strain gradient across thin GaAs waveguides offers an interesting playground for studying more elusive effects such as flexoelectricity \cite{yudin2013flexo} and its application in nano-electromechanical systems.

\section{Backmatter}

\begin{backmatter}
\bmsection{Funding}
We acknowledge funding from the European Research Council (ERC) under the European Union’s Horizon 2020 research and innovation program (No. 949043, NANOMEQ), the Danish National Research
Foundation (Center of Excellence “Hy-Q,” grant number DNRF139),Styrelsen for Forskning og Innovation (FI) (5072-
00016B QUANTECH), and the BMBF contract (No. 16KISQ009).

\bmsection{Disclosures}
The authors declare no conflicts of interest.

\bmsection{Data Availability Statement}
Data underlying the results presented in this paper are not publicly available at this time but may be obtained from the authors upon reasonable request.

\end{backmatter}
\bibliography{sample}

\end{document}